# Phase-Matched Generation of Coherent Soft-X-Rays

Andy Rundquist, Charles G. Durfee III, Zenghu Chang, Catherine Herne, Sterling Backus, Margaret M. Murnane, and Henry C. Kapteyn
Center for Ultrafast Optical Science
University of Michigan
Ann Arbor, MI 48109-2099
Ph. (313) 763-5574; FAX (313) 763-4876; E-mail: murnane@umich.edu
*To whom correspondence should be addressed. E-mail: murnane@umich.edu

**Abstract**

Phase-matched harmonic conversion of visible laser light into soft x-rays was demon- strated. The recently developed technique of guided-wave frequency conversion was used to upshift light from 800 nanometers to the range from 17 to 32 nanometers. This process increased the coherent x-ray output by factors of $10^2$ to $10^3$ compared to the non–phase-matched case. This source uses a small-scale (sub-millijoule) high repetition- rate laser and will enable a wide variety of new experimental investigations in linear and nonlinear x-ray science.

Nonlinear optical techniques for frequency conversion have played a pivotal role in the development of efficient coherent light sources (*1*), second in significance only to the invention of the laser itself. In recent years, nonlinear optics has led to the development of efficient visible-wavelength laser sources and sources with broad tunability such as optical parametric oscillators (*2*) and amplifiers (*3*). Most recently, the use of structured materials has made possible a new class of efficient nonlinear optical devices based on quasi–phase-matching (*4*).

However, nonlinear frequency-conversion techniques have almost exclusively re- lied on crystalline solids as nonlinear media. For efficient conversion, the electromagnetic waves corresponding to the driving and the generated signal must be phase-matched— both colors of light must have the same phase velocity as they travel through the generating medium. In this case, the nonlinear polarization adds coherently as the waves co-propagate, resulting in a rapid increase in signal intensity. To date, phase-matching has generally been achieved by relying on a combination of anisotropic materials and differing source and signal polarizations, because unless conditions are chosen carefully, light of different colors travel at different speeds through a material.

This reliance on solid materials has severely limited the application of nonlinear optical techniques to very short wavelengths. The best solid nonlinear materials such as LBO and BBO can be used to generate light at wavelengths as short as 200nm. However, few solids are transparent at wavelengths shorter than this, and none are transparent in the extreme ultraviolet (XUV) and soft x-ray regions of the spectrum. In contrast, many gases are transparent well into the vacuum ultraviolet (VUV), and even shorter wavelengths can propagate with moderate absorption through low-pressure gasses. However, because gases are isotropic, established phase-matching techniques are not applicable and thus phase-matched frequency conversion techniques have not been developed for the XUV and soft x-ray regions of the spectrum.

Nevertheless, in recent years the technique of high-harmonic generation has proven to be a useful source of XUV and soft x-ray light (6, 7). In this process, atoms in a dilute gas radiate harmonics of the incident light in the process of being ionized at the focus of an intense ultrashort-pulse laser. The highly nonlinear nature of the ionization process makes it possible to generate harmonics up to order 299 and higher, at wavelengths below 3 nm using very short laser pulses. (8, 9) Despite the low efficiency of this process  ($< 10^{-8}$ conversion of laser light into one harmonic peak), the photon flux generated is still sufficient to be useful for some applications in ultrafast x-ray spectroscopy. The high-harmonic generation (HHG) technique is also of considerable theoretical interest as a possible method for generating attosecond-duration light pulses (10, 11, 12).

However, many more applications would be possible if the process were made more efficient by phase-matching. Unfortunately, the rapidly changing free electron density associated with ionization of the atoms led to the generally-held opinion that phase-matching would result in at-most modest improvements in efficiency (13, 14, 15) because the length over which the harmonic phase slips by p from the laser (the coherence length) is ≈ 50μm for a fully ionized gas. Phase-matching would compensate for this phase slippage, as shown in the inset to Fig. 1, and allow the x-ray emission from much longer lengths to add coherently.

In this paper, we report a new phase-matching technique that circumvents the limitations of crystalline nonlinear optics by making phase-matched frequency conversion of ultrashort-pulse laser light into the XUV possible. Phase-matched frequency conversion in gases has been achieved in the UV in past experiments, by exploiting the negative dispersion which exists for light at wavelengths just shorter than a resonant absorption line (16, 17). However, the resonant requirement limits the phase-matching bandwidth, making it difficult to use this technique for applications which require broad tunability or ultrashort pulses, and also limits the conversion efficiency. Moreover, most experiments in high harmonic generation to date have been done in the tight focusing geometry, where the phase shift due to the laser focus dominates and limits phase-matching. Although proper choice of focusing conditions can somewhat increase the coherence length, the useful interaction length is still fundamentally limited by the laser beam divergence. In contrast, by propagating a light pulse through a waveguide, its phase velocity can be carefully controlled, and the beam divergence effects eliminated. Past work has proposed the use of a transient waveguide created by a laser-produced plasma for phase-matched frequency conversion using high-order frequency mixing. However, as will be shown below, a hollow capillary tube allows phase-matching in neutral gas with single frequency input, while employing a much simpler experimental configuration. The phase velocity of the laser light in a hollow waveguide can be adjusted by simply changing the guide diameter or by changing the gas pressure inside the waveguide. Thus, very long coherence and interaction lengths can be obtained, with a corresponding increase in conversion efficiency.

In past work, using the guided-wave frequency conversion technique of phase-matching, we demonstrated very high (20%) conversion efficiency of ultrashort light pulses into the uv region, using a four-wave mixing scheme. This finding immediately led to the question of whether guided-wave phase-matching could be applied to the x-rays generated as high-harmonic radiation. If high-harmonic generation is directly and inextricably linked to ionization, one must contend with the effects of a rapidly varying index of refraction. Furthermore, the intensities used for high-harmonic generation might cause damage to the waveguide structure. Even if the light did not cause damage, the reactive ionized gas might. In this work, we show that it is indeed possible to generate efficient high harmonic XUV radiation in a fiber, in a regime where ionization-induced index of refraction effects are minimized. We achieve this by using short (20fs) pulses to create the x-ray harmonics. A shorter pulse reduces the level of ionization at which a particular pulse intensity is reached, making it possible to generate high harmonics in a regime where the ionization-induced index does not preclude phase-matching. Experimentally, we have generated > 0.2 nJ of x-rays per harmonic per pulse at 1 kHz repetition-rate, with pulse durations expected to be under 5 fs, in a near-diffraction-limited beam. Using this source, it should be possible to produce focused x-ray intensities of greater than 1014 Wcm-2, assuming f/10 focusing. Straightforward improvements to our scheme should lead to further enhancements in x-ray output by at least an order of magnitude. Therefore, using higher repetition rate (10kHz) millijoule-level lasers, it should be possible to generate fractions of a mW power per harmonic peak. Even the current several µW's of power over 10 harmonic orders compares favorably with the average power produced by the current generation of soft-x-ray lasers. The development of our source will thus make possible new types of investigations in linear and nonlinear x-ray science and technology.

Guided-wave frequency conversion is possible because of the frequency-dependent dispersion introduced by light propagation in a waveguide. By confining the beam, a waveguide structure gives an additional, geometric component to the phase velocity. In a ray picture, the inside walls of the hollow waveguide confine the beam by repeated glancing-incidence reflections. The intrinsic diffraction, and thus the effective grazing angle of incidence on the walls, increases with wavelength. Because the projection of the wave vector along the optical axis is shorter than in free space, the phase velocity of the guided beam increases with wavelength. From another perspective, the discrete modes of the waveguide are standing waves in the transverse direction. When the waveguide k-vector, corresponding to the wavelength of this standing wave, is added to the free-space wave vector, diffraction is canceled. Because the waveguide k-vector is frequency independent, while the free-space propagation k-vector is longer for shorter wavelength, the net k-vector is most strongly perturbed for long-wavelength light. Mathematically, for light traveling in the wave guide, the k-vector of propagation is:

$$k \approx \frac{2\pi}{\lambda} + \frac{2\pi N_a \delta(\lambda)}{\lambda} - N_e r_e \lambda - \frac{u_{nm}^2 \lambda}{4\pi a^2}$$

where the first term corresponds to simple vacuum propagation, and the second and third result from dispersion of the gas and of the plasma created by ionization. Here, $N_a$ is the atom density, $N_e$ is the electron density, and d depends on the dispersive characteristics of the atom. The last term is due to the waveguide, where a is the radius of the wave guide, and $u_{nm}$ is a constant corresponding to the propagation mode in the fiber. The net phase mismatch for a nonlinear mixing process results from the vector sum of all of the waves involved, with source waves added and signal wave subtracted. Phase matching corresponds to $\Delta k = qk_{laser} - k_{x-ray} = 0$, where q is the harmonic order. The gas-filled hollow waveguide allows many adjustment parameters to engineer the phase matched-condition: the wavelength, gas pressure, gas species, waveguide size, and the spatial mode.

The case of high-harmonic generation is particularly simple. The interactions of the soft x-ray light with the waveguide is minimal, and the phase velocity of the x-rays is thus slightly greater than c (the speed of light in vacuum). The phase velocity of the laser is the balance of three competing terms: the waveguide dispersion and plasma dispersion which increase the phase velocity, and the material dispersion which decreases it. Because these contributions are of opposite sign, by varying the pressure of the gas, we can achieve the phase-matched condition where the x-ray and laser light travel with the same speed. This is shown conceptually in the inset to Fig. 1. The presence of ionization-induced plasma changes the dispersion substantially. However, by varying the gas pressure, phase-matching can be regained as long as the total percentage of ionized atoms is still small (less than ~5% for argon). The use of 20 fs pulses allows us to keep the total ionization fraction small but still achieve intensities that are sufficient for generating XUV harmonics of order up to 45.

A schematic of our experiment is shown in Fig. 1. We used a titanium-doped sapphire laser amplifier system, with a wavelength of 800 nm (1.55 eV photon energy). Pulses from a Kerr-lens mode locked oscillator were amplified at a repetition rate of 1 kHz in two multipass laser amplifier stages capable of generating 4.5 mJ in a 17 fs pulse duration. For these experiments, pulses of energy 100 to 300 µJ, with pulse duration 20 fs, were directed through an iris and focused with a 300 mm focal length fused silica lens through a sapphire chamber entrance window into a differentially pumped capillary cell. The laser waist size is 50 µm in the fiber. The capillary cell is a novel 3-segment design that allows a relatively high and more importantly constant gas pressure within the capillary, while greatly restricting the flow of gas into the vacuum regions at either end. If a pre-fiber were not used, the presence of gas before the first fiber could lead to ionization-induced defocusing of the laser beam. Three sections of capillary tube, with 150 μm inside diameter and 6.3 mm outside diameter, were held in a V-groove with ≈ 1 mm spacing between the segments. This design allowed a guided laser beam to pass from one section to another with minimal loss. The total capillary length was 6.4 cm, with a central section length of 3 cm. Flexible coupling to the vacuum system on either side permitted the capillary to be aligned properly. The input iris was used to adjust the focal spot diameter to match that of the lowest order mode of the capillary waveguide. The light emerging from the capillary was aligned through an imaging grazing incidence spectrometer (Hettrick Scientific HiREFS SXR-1.75). A thin aluminum filter (0.2 μm) was used to block the laser light and pass only the 11th through 45th harmonics (17 to 70 eV). The signal was detected with a microchannel plate coupled to a phosphor screen. (XSI Inc.). The image was either viewed with a charge-coupled device (CCD) camera or integrated with a photodetector. Laser damage to the capillary fiber was not a problem; although misalignment at high power can cause damage, with proper alignment and at the intensities used for these experiments, no degradation of the capillary waveguide was observed after tens of hours of operation at 1 kHz repetition-rate. The pumping speed required to maintain good vacuum in the spectrometer is also very greatly reduced compared with harmonic generation in a pulsed-gas nozzle or free-flowing cell, which have been used in the past for HHG experiments.

If phase-matching of conversion of laser light into x-rays is achieved, we should observe a dramatic increase in x-ray output at the optimum pressure for a given harmonic order. Propagation in the waveguide is basically plane-wave, and losses of the laser over the propagation length in the fiber (a few centimeters in these experiments) are minimal. Thus, for any given position temporally within the pulse and co-propagating with it, the level of ionization is constant - an optimum pressure and intensity can be found somewhere within the waveguide, either in the constant-density middle section, or in the varying-density end section. Typical data for the 29th and 31st harmonic of 800nm at 28 and 26 nm respectively, generated in a 150 μm capillary filled with argon, are shown in Fig. 2. The theoretically predicted output is shown in Fig. 3. At pressures of 35 torr, phase-matching is predicted and observed to occur and caused a dramatic increase in harmonic output. We checked that for pressures up to 200 torr, no change in the transmission of the laser light due to ionization effects occurred. Therefore, the decrease in signal at high pressures is due to the loss of phase matching because of excess gas dispersion, and not due to ionization-induced defocusing. A further confirmation of this explanation is that the optimum phase-matched pressure shifts to higher pressures if we increase the laser intensity. The increased intensity adds a negative contribution to the refractive index due to increased ionization, which must be balanced by a higher gas pressure. At pressures of 30 to 50 torr, a total of 5 harmonic peaks (orders 23-31) are phase matched. However, the optimum phase matching pressure for each harmonic is slightly different, because the higher harmonics are generated at higher values of ionization. No signal was observed for harmonic orders below the 23rd, because of strong absorption in the argon gas. This is a strong indication that the observed harmonic light is generated primarily within the central fiber segment. Also, at the peak intensities used in these experiments, ($\approx 2 \times 10^{14}$ Wcm$^{-2}$), we only expect to observe harmonics up to order 31. The higher harmonic orders (33 - 45) are phase-matched by increasing the gas pressure and laser intensity.

Figure 3 A to D shows the theoretically predicted output for the 29th harmonic, with and without the presence of absorption and ionization. The presence of absorption and ionization tends to dampen and broaden the expected oscillations of the output signal with pressure, both because the absorption depth (8 mm) is shorter than the coherence length, and because the harmonics are generated over a range of ionization levels, each optimizing at a slightly different pressure. The presence of ionization ($\approx$ 2% when the harmonic is being generated) tends to shift the optimum phase-matched pressure to higher values, as shown in Fig. 3B and D, because of the need to balance the refractive index contribution introduced by ionization to maintain phase-matching. From Fig. 3, we see that absorption in the gas reduces the expected output by an order of magnitude. This absorption explains why our output is enhanced only by factors of $10^2$ to $10^3$ (see below), instead of the maximum expected enhancements of $10^4$ due to the increase in coherence length from 50μm to an interaction length of a few centimeters. The coherence length is infinite when phase-matched.

The conversion efficiency of laser light into a single harmonic peak was measured in two independent ways. First, a vacuum photodiode (XSI Inc.) was inserted before the x-ray spectrometer, and the current was recorded with a picoammeter (Keithley 485). A metal filter was placed between the source and detector to block any low-order harmonics that might have been present. In practice, this procedure was not necessary because the low order harmonics are not phase matched at this pressure and also are strongly absorbed in argon, and therefore their signal levels are orders of magnitude below that of the five phase-matched harmonics. Signal levels of ~100pA were observed. Given our repetition rate of 1kHz, a detector quantum efficiency of at best 0.08, a measured filter transmission of 10%, and an input laser energy of 150μJ per pulse, we estimate that $2 \times 10^7$ photons per harmonic peak per pulse were emitted. This value corresponds to an energy of 0.2nJ per harmonic peak per pulse, or a conversion efficiency of $10^{-6}$ into a single harmonic peak - a 100 times increase over previously measured efficiencies. We also estimated the enhancement factors by comparing our output from the fiber with that from a gas jet. We estimate an increase in efficiency of $10^2$ to $10^3$ due to phase-matching. We note that the use of the vacuum photodiode x-ray detector significantly improved the accuracy of our conversion efficiency measurements, compared with those made using a microchannel plate detector.

An image of the phase-matched harmonic output was obtained by placing an imaging microchannel plate detector (XSI Inc.) before the spectrometer, at a distance of 0.68m from the output of the capillary. The observed harmonic beam images for harmonics 23 to 31 are shown in Fig. 4. Figure 4A shows the harmonic output beam at low pressures, which is not phase matched. Because in the absence of phase-matching, the harmonic generation is not associated with any one particular mode of the waveguide, the beam quality is poor. In contrast, Fig. 4B shows the output harmonic beam at a pressure of 35 torr, when phase matching is optimized. Because the fundamental waveguide mode is now phase-matched to create the x-ray output, the x-rays generated have excellent spatial properties. The grid pattern which is apparent on the image is due to the $\approx$ 27 lines/cm mesh on which the x-ray filter is mounted. The x-ray beam full-width-at-half-maximum diameter is $\approx$ 1 mm. Given our source distance of 0.68m, and the source size of radius $\approx$ 20 μm, the measured diffraction angle of ~1 mrad is consistent with a near-diffraction-limited x-ray output beam.

In future experiments, more optimal laser-fiber coupling, larger diameter fibers (which reduce absorption loss), lower absorption gases and using shorter pulses, should allow significant increases in the efficiency of the phase-matched harmonic conversion process. By using recent advances in laser technology, it should be straightforward to generate milliwatts of power per harmonic peak. It should also be possible to apply this technique to shorter wavelengths. This technique should thus enable a wide variety of new experimental investigations in linear and nonlinear x-ray science.

The authors gratefully acknowledge support for this work from the National Science Foundation. H. Kapteyn acknowledges support from an Alfred P. Sloan Foundation Fellowship.

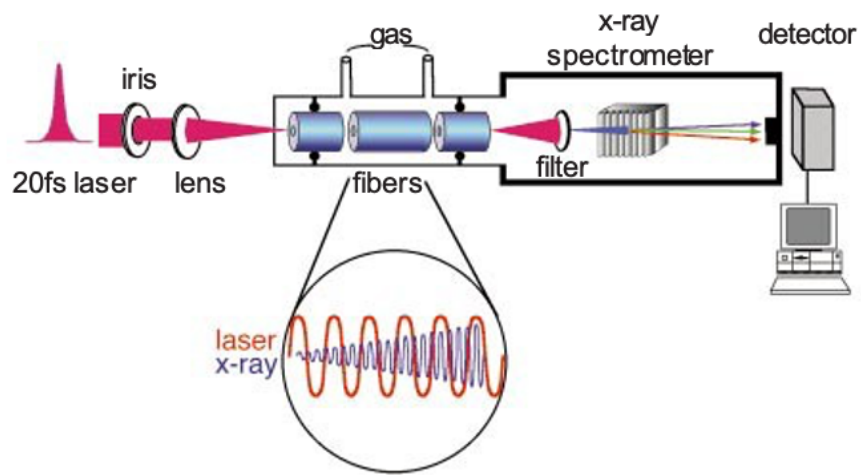

**Fig. 1.** Experimental set-up for phase-matching of soft x-rays in a capillary wave guide. The inset shows the growth of the x-ray wave when phase-matched. Note that the laser and x-ray wavelengths are not to scale.

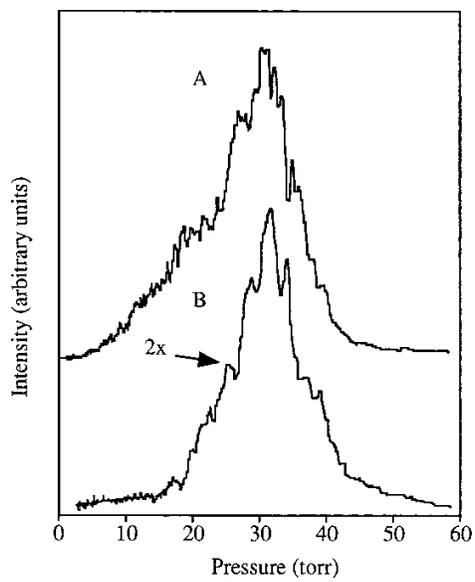

**Fig. 2.** Measured signals of the (A) 29th and (B) 31st harmonics as a function of gas pressure.

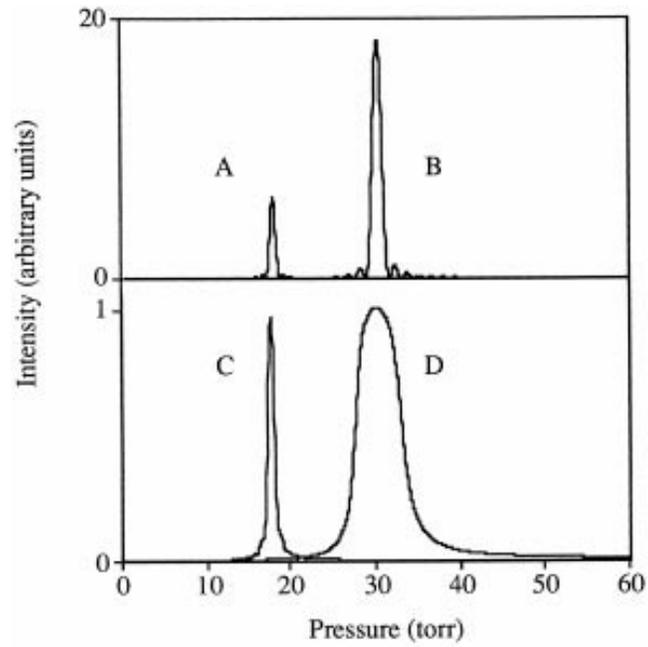

**Fig. 3.** (A) Calculated output of the 29th harmonic; (B) in the presence of constant 2% ionization; (C) in the presence of absorption; and (D) net output in the presence of absorption and varying levels of ionization around 2%.

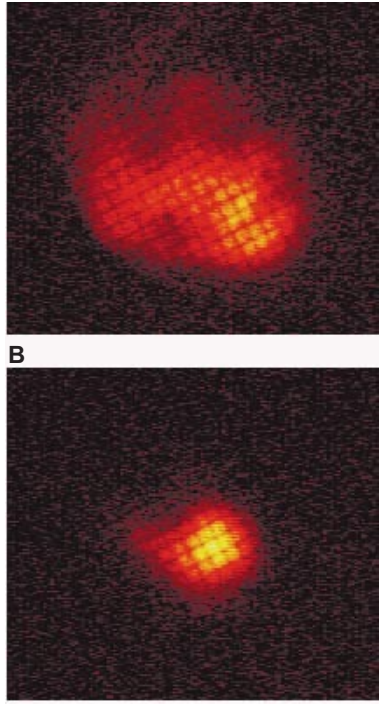

**Fig. 4.** Soft-x-ray output beam for the 23 - 31 harmonic orders: (A) un-phase-matched, detector set to high gain; (B) phase-matched, detector set to low gain.